\documentclass[oldversion]{aa}  
\usepackage{natbib}
\usepackage{graphicx}
\usepackage{txfonts}
\begin{document}
   \title{CO investigation of $z=0.4-1.5$ galaxies}

   \subtitle{}

   \author{A.-L. Melchior
          \inst{1,2}
          \and
          F. Combes\inst{1}
          }

   \offprints{A.L.Melchior@obspm.fr}

   \institute{LERMA,
Observatoire de Paris, LERMA, UMR8112, 61, avenue de l'Observatoire,
Paris, F-75014, France\\
              \email{A.L.Melchior@obspm.fr,Francoise.Combes@obspm.fr}
\and
Universit\'e Pierre et Marie Curie-Paris 6, 4, Place Jussieu,
F-75\,252 Paris Cedex 05, France\\ }

   \date{Received <date> / Accepted <date>}

 
  \abstract {We report on the results of an IRAM-30m search for CO
emission lines in three galaxies at intermediate redshifts. The idea
was to investigate the molecular content of galaxies bright in the
infrared at $z=0.4-1.5$, a redshift desert for molecular line studies,
poorly investigated as of yet. We integrated 8-10h per source and did
not succeed in detecting any of the sources. {From} our upper
limits, we are able to constrain the molecular gas content in these
systems to less than $4$ to $8\times 10^9 M_\odot$, assuming a
CO-to-H$_2$ conversion factor
($\alpha=0.8$~M$_\odot$/(K\,km\,s$^{-1}$\,pc$^{2}$)). We stress the
current difficulty of selecting sources with a detectable molecular
content, a problem that will be faced by the ALMA First Science
projects.}


   \maketitle
%

\section{Introduction}
Our current knowledge of the molecular content of galaxies at $z>0.4$
is currently limited to a fraction of the submillimetre bright objects
\citep{Solomon:2005,Greve:2005}, usually selected on the basis of
their strong infrared (IR) luminosity. The extent to which active
galactic nuclei (AGN) contribute to these extreme infrared luminosity
is currently {a} matter of debate. While the interpretation of the
IR-radio correlation in terms of on-going star formation is commonly
accepted \citep{Condon:1991,Condon:1992,Yun:2001}, this explanation
remains uncertain \citep[e.g.][]{Vlahakis:2007}.  The evidence that,
in the strongest infrared sources, a significant fraction of the
infrared luminosity is due to AGN \citep{Genzel:2000,Alexander:2005}
further complicates the interpretation of this
correlation. \cite{Farrah:2003} have shown a correlation of the AGN and
starburst luminosities over a wide range of IR luminosities.  The
recent detection of a molecular torus in Arp 220 \citep{Downes:2007}
demonstrates that the true source of at least part of its infrared
luminosity is due to a black hole accretion disc, while this galaxy
was considered as a prototypical starburst. Hence, there is the
possibility that the infrared luminosity is not a good tracer of star
formation activity, which is quite troublesome because this is
nevertheless the most reliable and easy tracer used so far (at least
unbiased by dust extinction).

In this complex context, the detection of molecular gas emission is
essential to get information about the mass and dynamics of these
galaxies, and to confirm the huge star formation rates
(750-1000\,M$_\odot$\,yr$^{-1}$) usually derived for (sub)millimetre
galaxies. Ultimately, this will contribute to further constrain the
scenario of hierarchical galaxy formation and evolution. However, the
current sensitivities are relatively low and require configurations
with huge amount of gas.  Different types of samples have been
investigated so far, with various success. On the one hand,
\citet{Evans:2006}, in their study of molecular gas in quasi-stellar
objects $z<0.15$, found a low CO to infrared luminosity ratio,
suggesting that the infrared luminosities of these QSO might be dominated
by the AGN component. \citet{Saripalli:2007} failed to detect any
molecular gas in restarting radio galaxies at $z<0.15$. On the other
hand, for the high-z galaxies with a CO detection, starbursts and AGN
do not exhibit significant differences in their molecular gas content
\citep{Solomon:2005,Greve:2005}. However, there is a clear difference
in the CO line widths, a factor of 2.3 narrower in velocity in QSO host
galaxies with respect to submillimetre galaxies, while powerful radio
galaxies fall in between \citep{Greve:2005,Carilli:2006}. The origin
of this effect is not yet clear: it could be due to a systematic in
the orientation of the QSO, but the possibility of merger signatures
in submillimetre galaxies is not excluded.

The redshift range $z=0.4-1.5$ corresponds to a key period of the star
formation history of the Universe: the end of the Star Formation
plateau \citep[e.g.][]{Madau:1998,Dahlen:2007} converging to the
nearby galaxy population. This range has been poorly investigated in
CO so far due to the lack of appropriate detectors and is known as a
redshift desert for molecular studies. This desert is due to the
presence of atmospheric lines (O$_2$), which prevent the
detection of CO(1-0) below 81\,GHz. (A similar situation was observed
in optical spectroscopy for galaxies with ${z=1.5-2.5}$
\citep{Steidel:2004}.) However, while most $z>1.5$ sources detected in CO are
magnified, {the} lower redshift range {is} relatively more
favourable to CO detection because there is no negative K-correction
for CO
\citep{Combes:1999}.

In this paper, we discuss the search for CO emission lines in three
galaxies at intermediate redshifts, in order to investigate the
molecular gas content of galaxies bright in the infrared at $z=0.4-1.5$.

Throughout this paper, we adopt a flat cosmology, with
$\Omega_\mathrm{m}=0.24$, $\Omega_\Lambda=0.76$ and
$H_0=73$~km\,s$^{-1}$\,Mpc$^{-1}$ \citep{Spergel:2007}.


\begin{table*}
 \centering
\caption{\protect Characteristics of the galaxies studied in this
paper. Listed from left to right are the names of the sources, their
J2000.0 positions, infrared luminosities, rest-frame B-band
luminosities, galaxy types, redshifts, and luminosity distances. The
appropriate references for the listed infrared luminosities and
redshifts are given in the rightmost column. The values of
L$^\mathrm{rest}_B$ are from \citet{Weiner:2005} for the CFRS sources,
and computed with the 2MASS J-band flux for NGP9\,F268-0341339.
}
\begin{tabular}{lllllllll} 
\hline
Source & RA (J2000) & DEC (J2000) & L$_{IR}$  ($10^{11}$\,L$_{\sun}$)&
L$^\mathrm{rest}_{B}$ ($10^{11}$\,L$_{\sun}$) $[$ L$_*^B$ $]^a$ & Type &
Redshift & $D_L$ (Gpc)& Ref. \\ \hline
\object{CFRS\,14.1329} & 14:17:34.8 & +52:27:52.0     & $1.3\pm 0.2$$^b$ & $0.04$ $[2]$& LIRG  &
0.375 &1.96  & 1, 2 \\ 
\object{CFRS\,14.1157} & 14:17:41.9 & +52:28:24.0    & $67\pm 4$ &$0.53$  $[25]$ & ULIRG & 1.149 & 7.84 &3, 2 \\ 
\object{NGP9\,F268-0341339} & 12:28:47.4 & +37:06:12.3 & $ -- $ & $6.9$ $[328]$ & QSO,
synch. & 1.515 & 11.12 & 4\\ \hline 
\end{tabular}
 \begin{minipage}{160mm}
1: \cite{Zheng:2004}; 
2: \cite{Hammer:1995}; 
3: LeFloch, private communication;
4:  \cite{Snellen:2002}. \\
$^a$ $ L_*^B = 2.1\times 10^9 L_{\sun}$ \citep{Marzke:1998}. $^b$
\citet{Flores:1999b} initially estimated L$_{IR} = $ 6.3$\pm$0.3$\times 10^{11} L_{\sun}$.
\end{minipage}
\label{tab:data}
\end{table*}

\section{Source selection}
We initially defined a sample of star-forming galaxies {from the}
Canada France Redshift Survey \citep[CFRS,][]{Lilly:1995}. We chose
galaxies with {an infrared-based} star formation tracer indicating
a SFR larger than 100 M$_\odot$\,yr$^{-1}$
\citep{Flores:1999b,Hammer:1995}. We also considered a subset of
strong radio-sources {from} the FIRST survey, for which redshifts
were available.  {In total,
three sources were observed: CFRS\,14.1329 and CFRS\,14.1157 (two
sources from the CFRS with ISO and VLA detections) and
NGP9\,F268-0341339 (a flat spectrum radio galaxy selected from the
FIRST survey).}\\ The infrared luminosities\footnote{Please note that
throughout the paper, L$_{IR}$ is defined as the integral of the flux
between 8 and 1000$\mu$m, while SFR$=1.71\times 10^{-10}
\frac{L_{IR}}{L_{\sun}}$ \citep{Kennicutt:1998}. Following
\citet{Elbaz:2002}, we took $L_{IR}=(1.91\pm0.17)\times L_{FIR}$,
where $L_{FIR}$ is defined in the range 40-120$\mu$m..}, available for
the CFRS sources \citep[][Le Floc'h, priv. comm.]{Zheng:2004}, are
provided in Table \ref{tab:data} together with the other (updated)
known characteristics of these galaxies. {Appendices \ref{sec:cfrs1}
and \ref{app:ngp9} provide more details about CFRS\,14.1329 and
NGP9\,F268-0341339.}

\section{CO Observations and reduction of the data}
We observed at IRAM-30m in May 2000 CO lines in the following
{galaxies:}
\object{CFRS\,14.1329}, \object{CFRS\,14.1157} and 
\object{NGP9~F268-0341339}. Table \ref{tab:data} displays their main
properties.  

Wobbler-switching mode was used, with reference positions offset in
azimuth by 90\arcsec\, for \object{CFRS\,14.1329} and
\object{CFRS\,14.1157} and 200\arcsec\, for \object{NGP9~F268-0341339}. At
1 and 3~mm, we used respectively 1~MHz filterbank and the
autocorrelator (1.25~MHz/channel) with bandwidths of 512 and 640~MHz.

The reduction was performed by the IRAM GILDAS
software\footnote{http://www.iram.fr/IRAMFR/GILDAS}.  For each line,
the spectra have been added and a polynomial of order 1 has been fitted
and subtracted. 
\begin{figure}
\centering
\includegraphics[width=8cm]{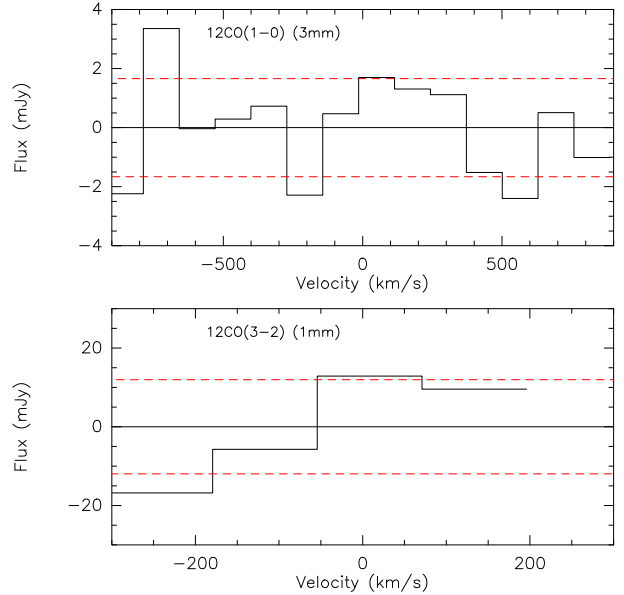}\\
\caption{Non-detection of the CO(1-0) and CO(3-2) lines searched in \object{CFRS\,14.1329} at
IRAM-30m (2000 May 3-4) at the spectroscopic redshift ($z=0.375$) of
the host galaxy. The displayed channel separations are
$128.7$~km\,s$^{-1}$ (top panel) and $125.2$~km\,s$^{-1}$ (bottom
panel). The dashed (red) lines indicate the 1-sigma levels reached for
this $8.5$~h integration.}
\label{fig:cfrs14.1329}   
\end{figure}
\begin{figure}
\centering
\includegraphics[width=8cm]{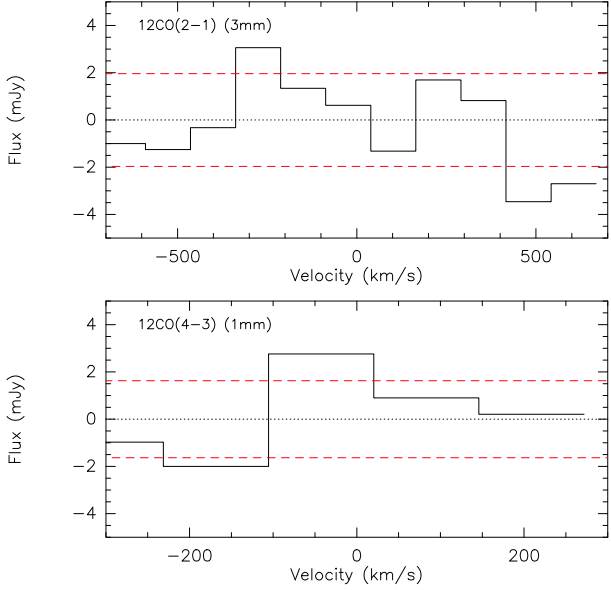}\\
\caption{Non-detection of the CO(1-0) and CO(3-2) lines searched in \object{CFRS\,14.1157} at
IRAM-30m (2000 May 4-5) at the spectroscopic redshift ($z=1.149$) of
the host galaxy. The displayed channel separations are
$125.8$~km\,s$^{-1}$ (top and bottom panel). The dashed (red) lines
indicate the 1-sigma levels reached for this $7.8$~h integration. }
\label{fig:cfrs14.1157}   
\end{figure}

\subsection{CFRS\,14.1329}
We searched for the CO(1-0) line at 83.83~GHz and the CO(3-2)
line at 251.49~GHz, relying on the spectroscopic redshift
$z=0.375$. At these frequencies, the telescope's half-power beam
widths are respectively 29\arcsec\, and 9.1\arcsec. We integrated
8.5~h on this source, with typical system temperatures of 203~K and
1117~K (on the T$^*_A$ scale). The observing conditions were not very
stable (wind).   We calibrated the spectra using the standard S/T$^*_A$ factors:
6.0 and 9.2 Jy\,K$^{-1}$.  As displayed in Fig. \ref{fig:cfrs14.1329},
we do not detect any line at the 1.7 and 12.0~mJy (rms) level.

\subsection{CFRS\,14.1157}
We searched for the CO(2-1) line at 107.28~GHz and the CO(4-3) line at
214.54~GHz, relying on the spectroscopic redshift $z=1.149$. At these
frequencies, the telescope's half-power beam widths are respectively
22\arcsec\, and 12\arcsec. We integrated 7.8~h on this source, with
typical system temperatures of 147.8~K and 324.1~K (on the T$^*_A$
scale).  We calibrate the spectra using the standard S/T$^*_A$
factors: 6.3 and 7.9 Jy\,K$^{-1}$.  As displayed in
Fig. \ref{fig:cfrs14.1157}, we do not detect any line at the 1.8 and
3.7 ~mJy (rms) level.  There is obviously some structures in the
baseline at 107.28~GHz.  However, this $1\sigma$ bump is too weak to
claim any detection and is most probably due to variable baselines, so
we do not apply any correction.

\subsection{NGP9~F268-0341339}
\begin{figure}
\centering
\includegraphics[width=8cm]{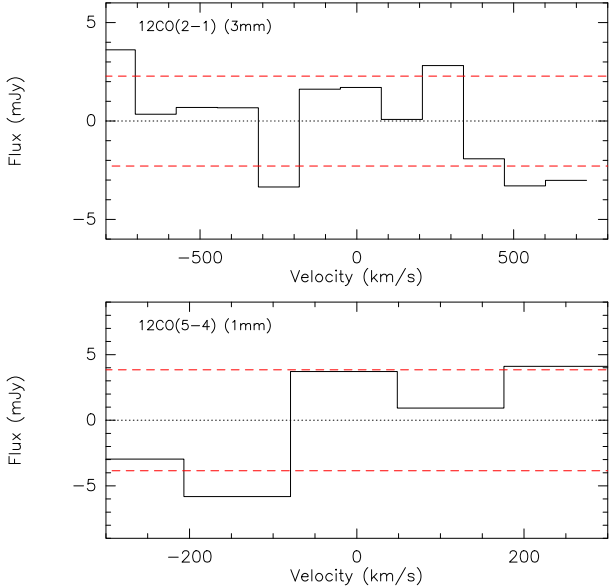}\\
\caption{Non-detection of the CO(2-1) and CO(5-4) lines searched in
\object{NGP9~F268-0341339} at IRAM-30m (2000 May 6-9) at the spectroscopic
redshift ($z=1.515$) of the galaxy. The displayed channel separations
are $130.8$~km\,s$^{-1}$ (top panel) and $127.6$~km\,s$^{-1}$ (bottom
panel). The dashed (red) lines indicate the 1-sigma levels reached for
this $10.35$~h integration.}
\label{fig:ngp9}   
\end{figure}
We searched for the CO(2-1) line at 91.670~GHz and the CO(5-4)
line at 229.130~GHz, relying on the spectroscopic redshift
$z=1.515$. At these frequencies, the telescope's half-power beam
widths are respectively 27\arcsec and 11\arcsec. We integrated
10.3~h on this source, with typical system temperatures of 147~K and
495~K (on the T$^*_A$ scale). The observing conditions were not very
stable.  We calibrated the spectra using the standard S/T$^*_A$
factors: 6.1 and 8.6 Jy\,K$^{-1}$.  As displayed in
Fig. \ref{fig:ngp9}, we do not detect any line at the 2.3 and 3.8~mJy
(rms) level.

\section[]{Analysis}
\label{sec:upper}
Given the secure spectroscopic optical redshift and the large
bandwidth at 3~mm, we do not expect a large velocity shift, which
could explain this missing CO emission. Very few galaxies (usually at
$z>3$) present a CO-line width (FWHM) larger than
1000~km\,s$^{-1}$. We would have detected a signal at 3~mm given our
reduction procedure.

{Following \citet{Seaquist:1995}, we calculate upper limits on the
velocity-integrated line fluxes using:}
\begin{equation}
 S_{{CO}} \Delta \mathrm{v} = 3 \sqrt{\Delta\mathrm{v_{\mathrm ch}}/
\Delta\mathrm{v}} \, \sigma_{\mathrm ch} \,\Delta\mathrm{v}
\end{equation}
where $\sigma_{\mathrm ch}$ is the channel-to-channel dispersion (rms)
computed in Jy for a given channel width $\Delta\mathrm{v_{\mathrm
ch}}$ (in km\,s$^{-1}$) and $\Delta\mathrm{v}$ is the (expected) line
width in km\,s$^{-1}$. {For the latter, we assumed a value of
300\,km\,s$^{-1}$.}  These upper limits are calculated from the final,
binned spectra shown in
Fig. \ref{fig:cfrs14.1329}, \ref{fig:cfrs14.1157} and \ref{fig:ngp9}.

{From the upper limits on the velocity-integrated line fluxes, the
corresponding constraints on CO line luminosities are computed as:}
\begin{equation}
L^\prime_{{CO}}= 3.25{\cdot} 10^{7} S_{{CO}} \Delta {\mathrm v}
\frac{D_{L}^2}{\nu_{{rest}}^2 (1+z)}
\end{equation}
where $L^\prime_{{CO}}$ is the CO-line luminosity expressed in
K\,km\,s$^{-1}$\,pc$^2$, $\nu_{{rest}}$ is the rest frequency of the
line in GHz, and $D_L$ the luminosity distance in Mpc \citep{Wright:2006}.

We expect that the CO flux ($S_{CO}$) is increasing as $\sim
\nu_{rest}^2$ for the first CO lines, for a given H$_2$ mass, as derived for
starbursts by Combes, Maoli $\&$ Omont \citeyearpar{Combes:1999}. The
ratios $L^\prime_{CO}(J=2-1)/L^\prime_{CO}(J=1-0)$,
$L^\prime_{CO}(J=3-2)/L^\prime_{CO}(J=1-0)$,
$L^\prime_{CO}(J=4-3)/L^\prime_{CO}(J=1-0)$ and
$L^\prime_{CO}(J=5-4)/L^\prime_{CO}(J=1-0)$ are thus taken to be equal
to 1. This assumes that the lines are thermalised at high temperature
and optically thick. For objects like quiescent nearby galaxies our
upper values should be multiplied by a factor up to 1.1
\citep{Braine:1992}, 1.6 \citep{Devereux:1994}, 2.2 and 4.8
\citep{Papadopoulos:2000}. However, our galaxies are probably starbursts not
representative of nearby sources, so we do not apply any correction.
\begin{table*}
 \centering
\caption{\protect Upper limits computed for the two  CO-lines
studied in each object (IRAM-30m observations). We provide $3\sigma$
upper limits on the line integrated intensity \protect $S_{CO}
\Delta$v and the CO line luminosity $L^\prime_{CO}$, assuming a line
width of 300\,km\,s$^{-1}$. }
\begin{tabular}{ll|ll|ll} 
\hline
{Lines} & {\object{CFRS\,14.1329}} &{Lines} &{\object{CFRS\,14.1157}} 
&{Lines} & {\object{NGP9~F268-0341339} }\\ \hline
&\multicolumn{5}{l}{Upper limits on \protect ${S_{CO}
\Delta \mathrm{v}}$~\protect(Jy\,km\,s${^{-1}}$)}\\ \hline
 CO(1-0) & $1.0$ & CO(2-1) & $1.05$ &CO(2-1) & $1.4$\\
 CO(3-2) & $7.0$ & CO(4-3) & $2.16$ &CO(5-4) & $2.2$ \\
\hline
& \multicolumn {5}{l}{Raw upper limits on ${L^\prime_{CO}~(10^9}$ K\,km\,s${^{-1}}$\,pc${^{2}}$) }\\ 
 \hline
 CO(1-0) & $6.9$ &CO(2-1) & $18.4$ & CO(2-1) & $41.1$ \\
 CO(3-2) & $5.3$ &CO(4-3) & $9.4$  & CO(5-4) & $10.7$ \\ \hline
\hline
\end{tabular}
\label{tab:upper}
\end{table*}

\begin{figure}
\centering
\includegraphics[width=8cm]{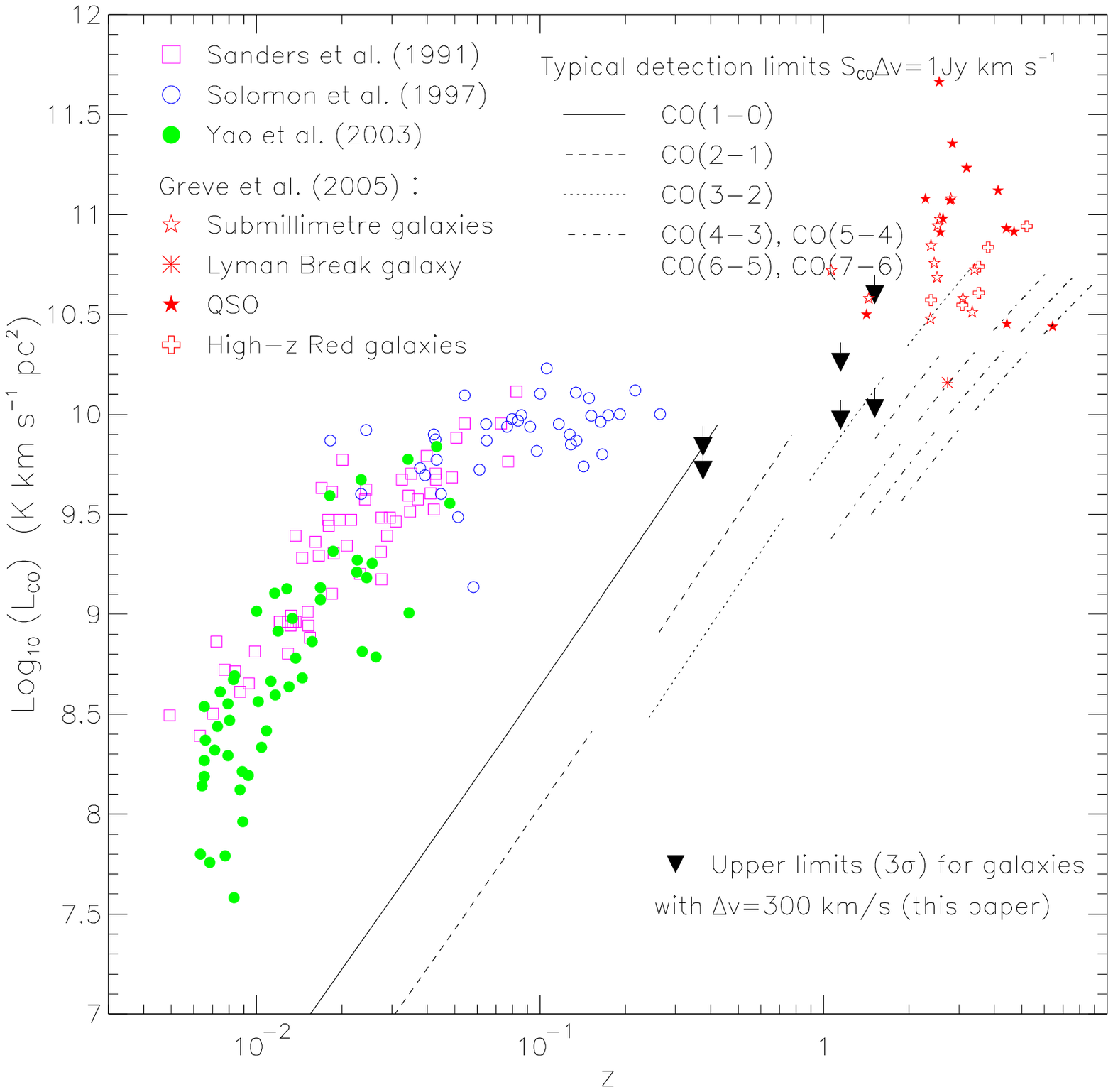}\\
\caption{CO-line luminosities \protect $L^\prime_{CO}$ as a function of
redshift. Upper limits (based on $\Delta$v$ = 300$\,km\,s$^{-1}$)
measured here for the studied galaxies are superimposed on previous
measurements from \protect\citet{Greve:2005}, \citet{Yao:2003},
\citet{Solomon:1997} and Sanders, Scoville $\&$ Soifer
\citeyearpar{Sanders:1991}. The two upper limits (3$\sigma$),
displayed for each source, correspond to the two measurements
performed at 1 and 3\,mm.  None of the CO-line luminosities has been
corrected for gravitational lensing. The curves correspond to an
integrated flux S$_{\mathrm CO} \Delta$v$=1$\,Jy\,km\,s$^{-1}$.}
\label{fig:plotgreve}
\end{figure}
Figure \ref{fig:plotgreve} displays the upper limits derived from our
observations on $L^\prime_{CO}$, compared to previous detections of
submillimetre galaxies detected in CO
\citep{Greve:2005,Yao:2003,Solomon:1997,Sanders:1991}. These
limits can be compared to the IRAM-30m best detection limits. They are
displayed in Fig. \ref{fig:plotgreve} and correspond to
S$_\mathrm{CO}\Delta{v} = 1$\,Jy\,km\,s$^{-1}$ for various CO lines
achieved at IRAM-30m. They illustrate the coverage of CO(1-0)
measurements and the complementarity of the other CO transition lines.

\begin{figure}
\centering
\includegraphics[width=8cm]{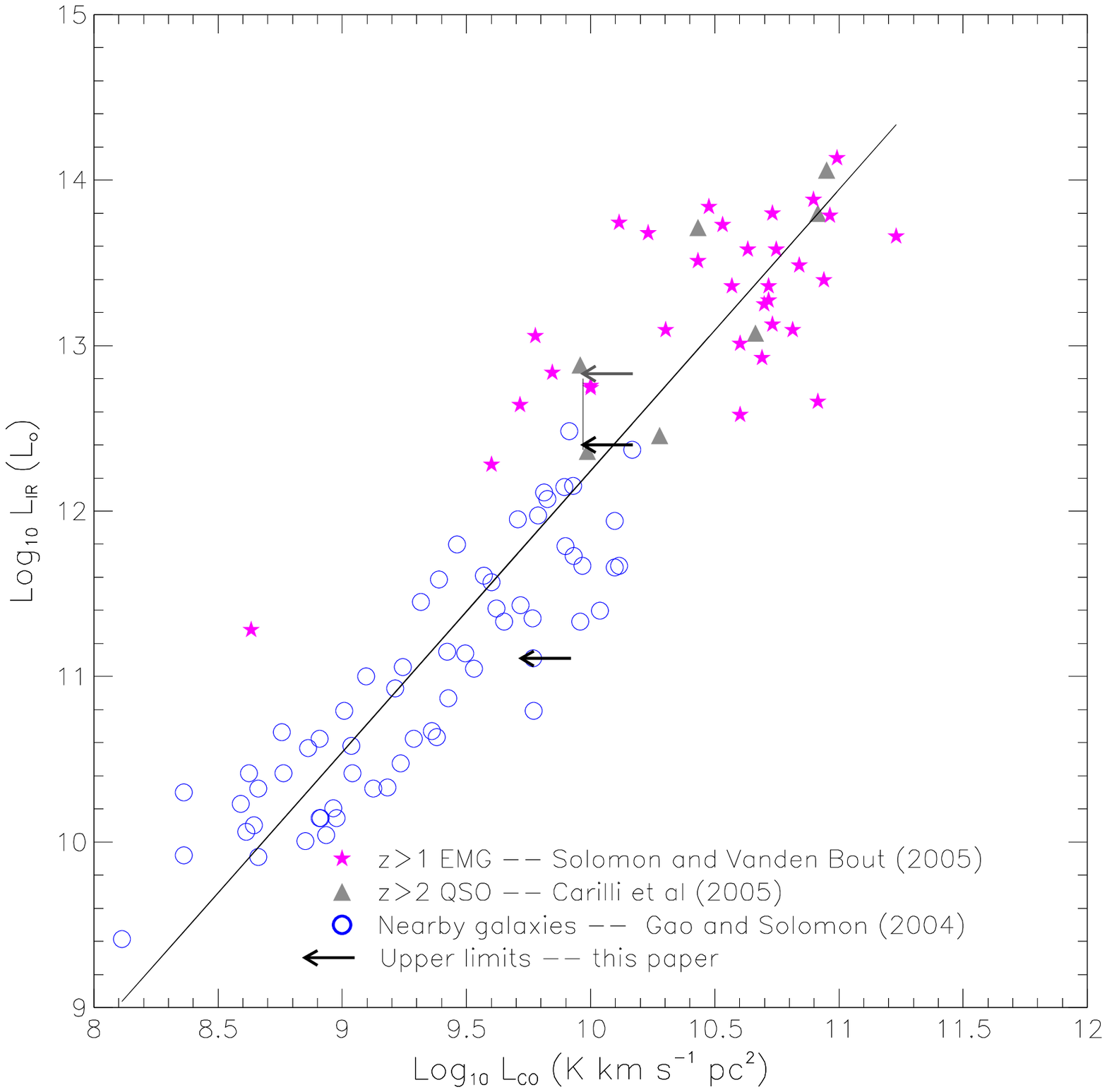}\\
\caption{The infrared and CO luminosities of nearby galaxies
\citep{Gao:2004}, ($z>2$) QSO \citep{Carilli:2005} and ($z>1$) Molecular
Emission Line Galaxies \citep{Solomon:2005} are compared to the 1-mm
upper limits obtained for the CFRS sources studied in this paper. For
CFRS\,14\,1157, the total IR luminosity and the starburst contribution
estimated by \citet{LeFloch:2007} are considered. The line corresponds
to the fit obtained by \protect\citet{Solomon:2005}. The high-z data
are corrected for gravitational lensing magnification when the factor
is known. }
\label{fig:plotgao04}
\end{figure}
Fig. \ref{fig:plotgao04} exhibits the correlation between the infrared
and CO luminosities. {The upper limits are consistent with the
IR-CO correlation given its significant scatter. The two arrows
provided for CFRS\,14.1157 correspond to the total IR luminosity and
to the sole contribution of the starburst, as discussed in
Sect. \ref{sect:dis}.}

\section{Discussion}
\label{sect:dis}
The non-CO detections discussed here reveal the difficulties of
selecting favourable candidates. The scatter of the correlation
between the IR and CO luminosities (see Fig. \ref{fig:plotgao04}) is
large and probably linked to the fact that the IR luminosities does
not trace the star formation activity only. In addition, it is
difficult to {obtain} IR luminosities, given the lack of all-sky
survey since IRAS. {The only solution has been here to rely on IR
luminosities derived from shorter wavelength measurements
\citep[namely 15$\mu$m][]{Flores:1999b,Zheng:2004}.}

{The lack of CO detections towards \object{CFRS\,14.1329} and
\object{NGP9~F268-0341339} is probably due to the fact that they are
not as good candidates as we initially thought.} As explained in
Appendix \ref{sec:cfrs1}, \object{CFRS\,14.1329} had been selected on
the basis of a favourable $L_{IR}$ luminosity
\citep{Flores:1999b}, which was subsequently revised
\citep{Zheng:2004}. We also discussed in Appendix
\ref{app:ngp9} that \object{NGP9~F268-0341339} is behind a foreground galaxy at
$z=0.138$ with SFR$\sim 95\,M_{\sun}$\,yr$^{-1}$.

{From our upper limits on the CO(1-0) and CO(3-2) luminosities, we
estimate that the molecular gas content\footnote{Following
\citet{Solomon:2005}, our molecular gas content estimates
account for the He mass.} of \object{CFRS\,14.1329} is smaller than
$5.5\times 10^9 M_{\sun}$ and $4.2\times 10^9 M_{\sun}$, assuming a
molecular gas mass to CO luminosity ratio
$\alpha=0.8\,M_{\sun}\,($K\,km\,s$^{-1}$pc$^2)^{-1}$, typical of
{ultraluminous infrared galaxies (ULIRG)}
\citep[e.g.][]{Solomon:2005}. These total molecular gas masses are
consistent with the gas content of local normal spiral galaxies
\citep[]{Gao:2004}, while the infrared luminosity (derived from
15\,$\mu$m data) was suggesting a luminous infrared galaxy (LIRG).
Hence, the value of $\alpha$ is uncertain and probably underestimated.
For the more distant galaxy \object{NGP9~F268-0341339}, the limits are
less stringent and it is still compatible with an (U)LIRG. With the
same assumptions, the molecular mass is smaller than $33\times 10^9
M_{\sun}$ and $8.6\times 10^9 M_{\sun}$ according to our upper limits
on the CO(2-1) and CO(5-4) luminosities.  The luminosities derived
from higher CO line transitions are more stringent than those derived
from the lower ones as we assumed that the lines were optically thick
(see also Sect.  \ref{sec:upper}).}

In contrast, \object{CFRS\,14.1157} was a very favourable candidate
that has remained as such. \citet{Zheng:2004} estimated $L_{IR}=178.6
\times 10^{11}\, L_{\sun}$ (SFR$\sim 3054$\,M$_{\sun}$\,yr$^{-1}$),
relying on the IR-15$\mu$m luminosities correlation measured by
\citet{Elbaz:2002}. \citet{LeFloch:2007} published a panchromatic
spectral energy distribution of this galaxy with a very good 
wavelength coverage. They derived a direct estimate\footnote{Please
note that \citet{LeFloch:2007} worked with the 1-1000 $\mu$m range. In
this paper, we adapted their figures to the 8-1000$\mu$m range used
here, thanks to the corresponding value kindly provided by E. Le
Floc'h.} of the infrared luminosity computed over the range
8-1000$\mu$m : $L_{IR}=67\pm 4\,\times 10^{11} L_{\sun}$, which would
correspond to SFR$\sim 1150$\,M$_{\sun}$\,yr$^{-1}$. However,
\citet{LeFloch:2007} have estimated that 67.2$\%$ percent of the
infrared flux is due to the AGN component. Accordingly, the SFR is
probably of order 375\,M$_{\sun}$\,yr$^{-1}$. 

{From our upper limits on the CO(2-1) and CO(4-3) luminosities of
\object{CFRS\,14.1157}, we find that the molecular gas content of this
galaxy is smaller than $15\times 10^9 M_{\sun}$ and $7.5\times 10^9
M_{\sun}$} (with
$\alpha=0.8\,M_{\sun}\,($K\,km\,s$^{-1}$pc$^2)^{-1}$).  This supports
the idea that the observed infrared luminosity is probably dominated
by the AGN component, as estimated by \citet{LeFloch:2007}. In
addition, one can note that the SFR derived on the basis of the
$[$\ion{O}{II}$]$ line is significantly lower: $2M_{\sun}$\,yr$^{-1}$
relying on the equivalent width and rest-frame absolute magnitude
measured by
\citet{Weiner:2005} and the formula of
\citet{Guzman:1997}, while we estimate $3.5M_{\sun}$\,yr$^{-1}$ with
an integration of $[$\ion{O}{II}$]$ the spectra published by
\citet{LeFloch:2007} (see their figure 2a) and normalised to SDSS
fluxes. This could be compatible with the IR-derived SFR only with a
factor of extinction of $105-190$, while $L_{IR}/L_B=41$ if one
assumes that only one third of the infrared luminosity contributes to
the SFR. {While a large scatter is known to affect
$[$\ion{O}{II}$]$ luminosities as discussed by \citet{Weiner:2007},
the previous comparison stresses the importance of the actual fraction
of the infrared luminosity due to the starburst: this fraction is
often overestimated.}

\begin{acknowledgements}
We thank the DEEP collaboration for providing us with the optical
spectra of \object{CFRS\,14.1157}. We are most grateful to C. Willmer
and E. Le Floc'h, who help us in this procedure. This research has
made use of the NASA/IPAC Extragalactic Database (NED), which is
operated by the Jet Propulsion Laboratory, California Institute of
Technology, under contract with the National Aeronautics and Space
Administration.
\end{acknowledgements}

\appendix
\section{CFRS\,14.1329}
\label{sec:cfrs1}
This galaxy was first detected in radio by \citet{Fomalont:1991}. It
was subsequently detected in the CFRS survey
\citep{Lilly:1995,Hammer:1995}. It has been classified as a dusty Sa
by \citet{vandenBergh:2001}, while \citet{Flores:1999b} classified it
as a strong starburst and highly reddened starburst with a lenticular
morphology. \citet{Flores:1999b} fitted \citet{Schmitt:1997} {spectral
energy distribution} templates to optical, K and mid-infrared data,
{as well as to the} 60$\mu$m {flux} (derived from the
radio-FIR correlation). {The infrared luminosity\footnote{This
luminosity corresponds to SFR$=107\pm 5\,M_{\sun}\,$yr$^{-1}$ relying
on \citet{Kennicutt:1998}.} derived for \object{CFRS\,14.1329} in this
way was} $62.8\pm3.1\times 10^{10} L_{\sun}$.  We relied on this
infrared luminosity to select \object{CFRS\,14.1329} for CO
observations. In the meantime,
\citet{Zheng:2004} reanalysed a CFRS sample containing \object{CFRS\,14.1329}
with the mid-infrared and infrared correlation found by
\citet{Elbaz:2002}, and published a revised version of the infrared
luminosity a factor of $4.8$ smaller: $L_{IR}=13.0\pm 2.5 \times
10^{10}\,L_{\sun}$ (SFR$=22.3\pm 4.2\,M_{\sun}\,$yr$^{-1}$). This
revision reflects the uncertainties related to the determination of
the infrared luminosity based on correlations with other wavelengths
luminosities. This lower value is in agreement with our non-CO
detection (see Table \ref{tab:upper}). More recently, it has been
further observed with the DEEP Groth Strip Survey
\citep{Vogt:2005}. In addition, \citet{Hammer:2005} estimate
SFR($[$\ion{O}{II}$]$)=3.5\,M$_{\sun}$\,yr$^{-1}$, which would require
a factor of extinction of 6.4, while $L_{IR}/L_B=32.5$.

\section{NGP9~F268-0341339}
\label{app:ngp9}
This object is a radio-flat spectrum galaxy. Its optical spectra is
typical of an AGN. The SCANPI/IRAS tool suggests a possible signal at
12~$\mu$m. However, there is foreground disc galaxy
(\object{SDSS\,J122847.72+370606.9}) at $z=0.138$, which is hosting an
intense star formation activity (SFR$\sim 95$M$_{\sun}$yr$^{-1}$). Its
optical spectra is typical of an Sb galaxy, while it exhibits a strong
H$\alpha$ emission line (SFR$\sim 95$M$_{\sun}$yr$^{-1}$). Given its
close angular distance (5\arcsec) to \object{NGP9~F268-0341339}, it
most probably dominates the possible infrared IRAS fluxes.
\end{document}